\documentclass[
reprint,
superscriptaddress,
amsmath,amssymb,
aps,
prb,
longbibliography
]{revtex4-1}

\usepackage{graphicx,epsf}
\usepackage{float}
\usepackage{xcolor}
\usepackage{bm}

\usepackage{cancel}

\begin{document}

\title{Conductance of electrostatic wire junctions in bilayer graphene}

\author{Sungguen Ryu}
\affiliation{Institute for Cross-Disciplinary Physics and Complex Systems IFISC 
(CSIC-UIB), E-07122 Palma, Spain} 
\author{Rosa L\'opez}
\affiliation{Institute for Cross-Disciplinary Physics and Complex Systems IFISC 
(CSIC-UIB), E-07122 Palma, Spain} 
\affiliation{Department of Physics, University of the Balearic Islands, 
E-07122 Palma, Spain}
\author{Lloren\c{c} Serra}
\affiliation{Institute for Cross-Disciplinary Physics and Complex Systems IFISC 
(CSIC-UIB), E-07122 Palma, Spain} 
\affiliation{Department of Physics, University of the Balearic Islands, 
E-07122 Palma, Spain}

\begin{abstract}
The conductance of electrostatic wire junctions in bilayer graphene,
classified as trivial-trivial or trivial-topological regarding the 
confinement character on each junction side, is calculated.  The topological side always corresponds to a kink-antikink system, as required for a proper connection with a trivial side. We report a conductance quench of the trivial-topological junction, with a  conductance 
{\it near} quantization to $4e^2/h$, which is only half of the maximum value 
allowed by the Chern number of a  kink-antikink system.
The analysis allowed us to uncover the existence of a chiral edge mode
in the trivial wire under quite general conditions.
A double junction, trivial-topological-trivial, displays
periodic
Fano-like conductance resonances (dips or peaks)
induced by the created topological loop.
\end{abstract}

\maketitle

\section{Introduction}
\label{Intro}

Electrostatic confinement in bilayer graphene (BLG) induced by microelectrodes acting at a distance from the top and bottom sides of the graphene planes has attracted a
notable attention in the graphene community. \cite{Mcan13,Zhang13,rozhkov16,Over18,Kraf18,Eich18,Kurzmann19,Banszerus20,Banszerus21}
The physical principle behind this electrostatic confinement is
a spatial modulation of the asymmetry potential $V_a$ between the two graphene planes. Bulk BLG 
in the absence of 
asymmetry potential is gapless, while it becomes gapped around zero energy in presence of the asymmetry potential. The gap is proportional to $|V_a|$ and so the spatial 
modulation achieved with microelectrodes is able to create regions of confinement 
whose shape and size can be controlled by the geometry of the 
fabricated microelectrodes. 

Smooth electrostatic confinement 
in BLG avoids difficulties introduced by atomically rough 
edges made when parts of the graphene system are physically etched, such  as strong intervalley scattering induced by edge roughness.\cite{Men12}
Confinement by etching in graphene is
challenging mostly due to the resulting edge imperfections.
See, however, Ref.\ \onlinecite{Cle19} for a recent
experiment of monolayer graphene nanoconstrictions with low edge roughness showing conductance quantization in $2e^2/h$ steps.
As discussed below, we restrict in this work to BLG electrostatic confinement, where conductance quantization steps are $4e^2/h$
due to the valley degeneracy in absence of magnetic field.

\begin{figure}[b]
\begin{center}
\includegraphics[width=0.6\textwidth, trim = 6cm 6cm 2cm 2cm, clip]{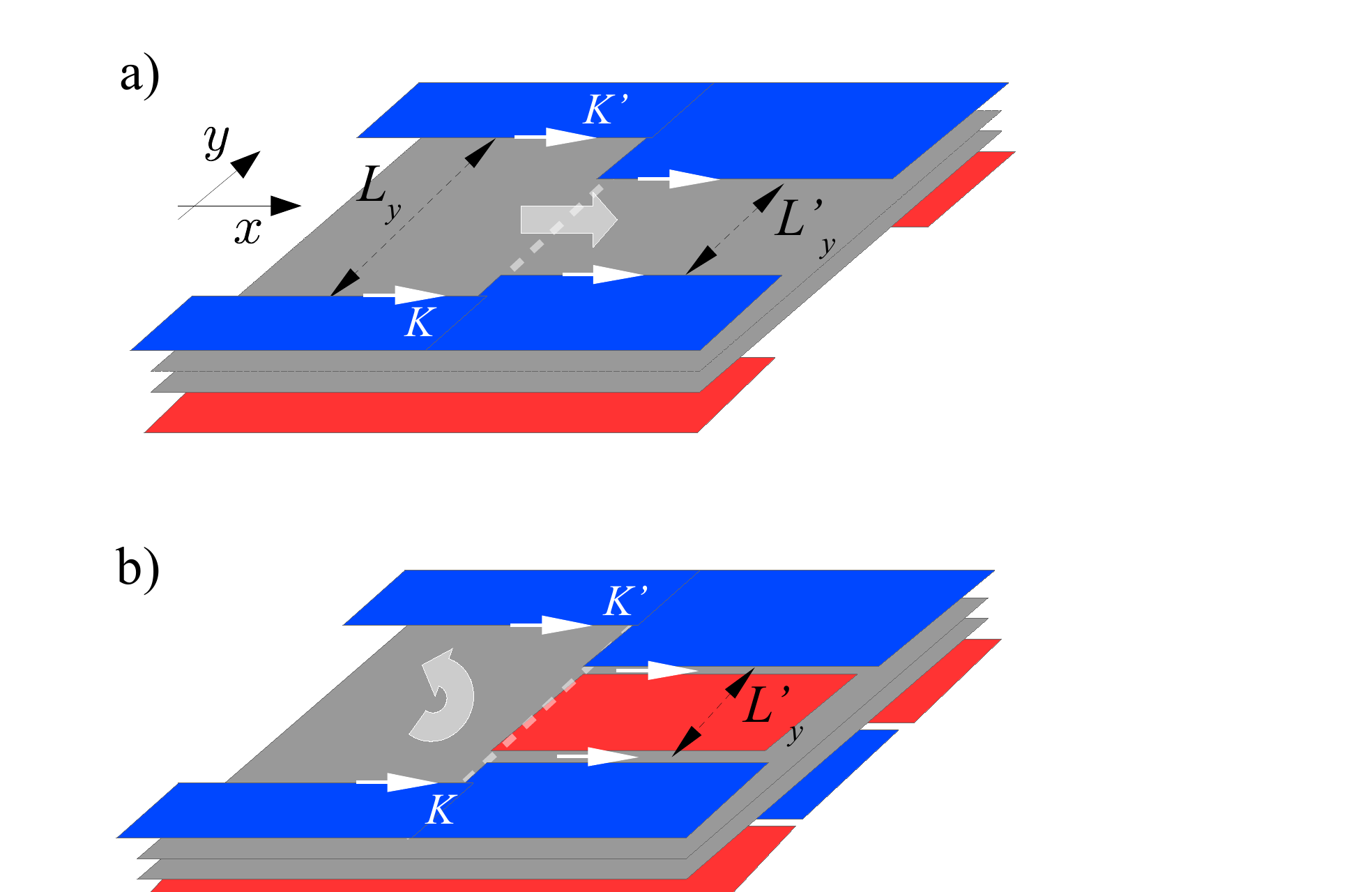}
\end{center}
\caption{Sketches representing BLG electrostatic junctions of type a) trivial-trivial and b) 
trivial-topological. The graphene layers are the 
grey planes and the top/bottom microelectrodes are the colored planes. 
Blue and red are indicating different signs of the applied potential
on the corresponding microelectrode.
The white dashed line is indicating the junction position. The
lateral widths on each junction side are $L_y$ and $L'_y$.
Bulk and edge electronic modes incident from the left are indicated by thick and thin white arrows, respectively. Notice that in b), bulk modes are mostly  backscattered
and only the edge chiral modes discussed in this work are transmitted
from left to right.
}
\label{Fig1}
\end{figure}

Two qualitatively different types of electrostatic confinement in BLG can be considered: a) {\it trivial} confinement when all top gates have the same potential,  which is opposed to that of all bottom gates,\cite{Pereira07,Recher09,Zarenia09,Pereira09,Zarenia10,Zarenia10b,daCosta14}
and 
b) {\it topological} confinement when the polarities of the microelectrodes are such that there are borders separating regions  
of opposite signs.\cite{Mar08,Zarenia11,xavier10,Ben21,Ben21b}
Figure \ref{Fig1} illustrates these two different
possibilities. Particularly, the topological 
confinement along two parallel lines of sign inversion (a kink-antikink system)
can be
seen on the right side of Fig.\ \ref{Fig1}b.
Both types of BLG confinement, trivial and topological,  have been 
intensively investigated theoretically, 
\cite{Pereira07,Recher09,Zarenia09,Pereira09,Zarenia10,Zarenia10b,daCosta14,Mar08,Zarenia11,xavier10,Ben21,Ben21b}
and importantly, also realized in experiments.\cite{Lon15,Men15,Li16,Eich18,Kurzmann19,Banszerus20,Chen20,Banszerus21}

The BLG trivial confinement has some similarities with 2DEG
confinement achieved in 
semiconductor nanostructures by modulated gating. In both systems, the quantum states are characterized by a sizeable region of 2D character; e.g., the inner part of quantum dots and quantum wires 
is 2D-like while its surrounding part shows an exponential decay across the border in the outer direction with respect to the bulk.  
The topological confinement, on the other hand, is characterized by a
predominantly 1D character; the wave functions 
vanishing with increasing distance in both directions from the border,
without requiring any 2D bulk.

This work focuses on junctions between electrostatically confined 
wires in BLG. Sketches of the junctions can be seen in Fig.\ \ref{Fig1}.  
We are particularly interested in the measurable differences between 
trivial-trivial and trivial-topological junctions that could be 
used to unambigously identify each confinement character.
We always consider a trivial side since, 
in practice, asymptotic leads are more likely to be trivial 
due to the above mentioned 2D character of this type of confinement.
The case of purely topological confinement, with scattering due to 
kink-antikink constrictions, was studied in Ref.\ \onlinecite{Ben21}.
After the single junction, 
this work also addresses the double junction with left and right 
trivial leads, highlighting the new features induced by a finite 
central region that may be trivial or topological.

Our main finding is a conductance quench of the trivial-topological 
junction (Fig.\ \ref{Fig1}b) as compared to the 
value given by the
number of propagating modes or Chern number of the topological 
side  ${\cal N}'_{\it top}$. 
While it is ${\cal N}'_{\it top}=2$ for a kink-antinkink, 
with additional valley and spin degeneracies, 
the junction conductance remains {\em nearly quantized} to 
$G\approx G_0$, where $G_0=4e^2/h$, 
instead of the {\it a priori} possible $G={\cal N}'_{top} G_0$.
That is, as a function of energy a large plateau with $G\approx G_0$
is found, even when the number of incident modes from the trivial side
increases up to ${\cal N}_{\it tri}\approx 4$.
In sharp contrast, the conductance of the trivial-trivial junction (Fig.\ \ref{Fig1}a) closely follows the Chern number of the right 
wire ${\cal N}'_{\it tri}$, 
showing a smooth staircase quantization $G={\cal N}'_{\it tri}G_0$, with 
${\cal N}'_{\it tri}=1,2,3,\dots$.

We explain the conductance quench of the trivial-topological junction by
realizing that only the lowest trivial mode (per valley and spin) is effectively transmitted 
from left to right, while the other modes are mostly reflected. 
Notice that there is a degeneracy factor 4, due to the accumulated valley and spin degeneracies
such that the corresponding conductance is the above defined $G_0\equiv4e^2/h$. As shown below, 
we found that 
the lowest mode
of a trivial wire acquires a remarkable chiral edge character for 
increasingly large magnitude of momentum $k$. 
Such property explains the nearly perfect transmission 
of the chiral edge mode on the left
to the 
topological chiral modes on the right side of the trivial-topological junction. The possibility of boundary modes in gapped BLG have been discussed in 
general terms in Ref.\ \onlinecite{Zhang13}.

Next, the work addresses the double junction systems. In the trivial-topological-trivial double junction closed loops can be formed 
in the central part for specific energies. We find that these closed-loop states yield 
conspicuous quasi periodic 
resonances, conductance peaks or dips of Fano type, as a function of energy or length of the central part.
Overall, our work suggests the conductance quench of the single junction and the periodic 
Fano resonances of double junctions as characteristic features 
signalling topological 
confinement in BLG electrostatic wire junctions. 

\section{Model and method}
\label{MM}

Our modelling of BLG nanostructures in based on the low-energy 
multiband Hamiltonian with continuum space operators for 
position $(x,y)$ and momenta $(p_x,p_y)$, as well as three 
pseudospin vectors, $\vec{\sigma}$, $\vec{\tau}$ and  $\vec{\lambda}$
for sublattice, valley and layer, respectively.\cite{Mcan13} 
In detail, the Hamiltonian reads
\begin{eqnarray}
 H &=& v_F p_x \tau_z \sigma_x
 + v_F\, 
p_y \sigma_y \nonumber\\
 &+& \frac{t}{2}\, \left(\,\lambda_x \sigma_x +\lambda_y\sigma_y\,\right) 
 + V_a(x,y)\, \lambda_z\; ,
\label{eq1}
 \end{eqnarray}
 where  
$\hbar v_F= 660\, {\rm meV}\,{\rm nm}$  and $t=380\,{\rm meV}$ are the 
BLG Fermi velocity and interlayer coupling, respectively.
We stress here that our notation of using different symbols for Pauli matrices 
in different subspaces while being more compact is also equivalent to other approaches using always the same symbols 
for Pauli matrices, irrespectively of the subspace, but keeping track of the strict
ordering of operators to define generalized matrices.\cite{Sny07,Gonz10} 
For example, 
for an operator like $\sigma_x\lambda_x$ we have the following equivalences
\begin{eqnarray}
\label{eqord}
\sigma_x\lambda_x
=
\lambda_x\sigma_x
&\Leftrightarrow&
\sigma_x^{{\it sublattice}}\otimes\sigma_x^{{\it layer}}\otimes
{\mathbf 1}^{{\it valley}}\nonumber\\
&\Leftrightarrow&
\left(
\begin{array}{cccc}
 0 & 0 & 0 & 1 \\
 0 & 0 & 1 & 0 \\
 0 & 1 & 0 & 0 \\
 1 & 0 & 0 & 0 \\
\end{array}
\right)
\otimes
{\mathbf 1}^{{\it valley}}\; .
\end{eqnarray}
Symbol $\otimes$ is used to indicate tensor product 
of different subspaces and
the last equivalence in Eq.\ (\ref{eqord}) assumes a specific 
spinor ordering
$(A1,B1,A2,B2)$, with $A/B$ and $1/2$ indicating sublattice and 
layer, respectively.

The position dependent asymmetry potential $V_a(x,y)$
in Eq.\ (\ref{eq1})
is chosen according to the distributions of microelectrodes, as shown in Fig.\ \ref{Fig1}. We assume the saturating potential values $+20\,{\rm meV}$  $(-20\,{\rm meV}$) on the graphene planes beneath the blue (red) colored microelectrodes, with smooth transitions at interfaces 
of a diffusivity $s=12$ nm. The smooth asymmetry potentials are modeled with logistic functions.\cite{Ben21}
$V_a(x,y)$ is piecewise defined
depending on the position $x$ either 
as trivial $V_a^{\it triv}(y)$ or topological 
$V_a^{\it top}(y)$,  
as indicated in Fig.\ \ref{Fig1}.
Assuming transverse boundaries at 
$y_a$ and $y_b$ $(>y_a)$
the potentials read
\begin{eqnarray}
V_a^{\it triv}(y) &=&
V_a\, \left(
1 
+\frac{1}{1+e^{\frac{y-y_a}{s}}}
-\frac{1}{1+e^{\frac{y-y_b}{s}}}
\right)\; ,
\\
V_a^{\it top}(y) &=&
V_a\, \left(
1 
+\frac{2}{1+e^{\frac{y-y_a}{s}}}
-\frac{2}{1+e^{\frac{y-y_b}{s}}}
\right)
\; .
\end{eqnarray}
Notice that the asymmetry potential vanishes in the 2D-bulk region of trivial confinement, in the central part ($y_a<y<y_b$) of the trivial wires.

Our junction modelling is based on complex-band structure theory.\cite{Osca19} 
The method proceeds in two steps;
first, a large set of complex-$k$ eigenmodes is determined in each piece of a junction by matrix diagonalization; second, a system of linear equations 
describing the wave-function matching at the junction interface is solved for each incidence condition.
The conductance is obtained from the transmissions $t_{kk'}$ with Landauer's formula 
$G=G_0\sum_{kk'}{|t_{kk'}|^2}$.
High numerical efficiency is achieved by exploiting the sparse character of the matrix which is diagonalized 
and of the matrix for the matching condition at the junction interface. Details of the method were presented in Appendix C of Ref.\ \onlinecite{Ben21}.
An important aspect is the proper filtering of spurious solutions that 
emerge due to Fermion doubling.\cite{Susskind77,Nielsen81,Lewe12}
In our complex band structure
approach only a 1D $y$-grid is required at the junction interface. This allows a high spatial resolution and thus a good filtering of the spurious
states by means of a coarse graining. We refer the reader to Refs.\ \onlinecite{Osca19,Ben21} for more details on the modelling method, focussing
next on the specific physical results.

\section{Results and discussion}
\label{s3}

\subsection{Wire single junctions}
\label{s3a}

\begin{figure}[t]
\begin{center}
\includegraphics[width=0.5\textwidth, trim = 1cm 6cm 1cm 3cm, clip]{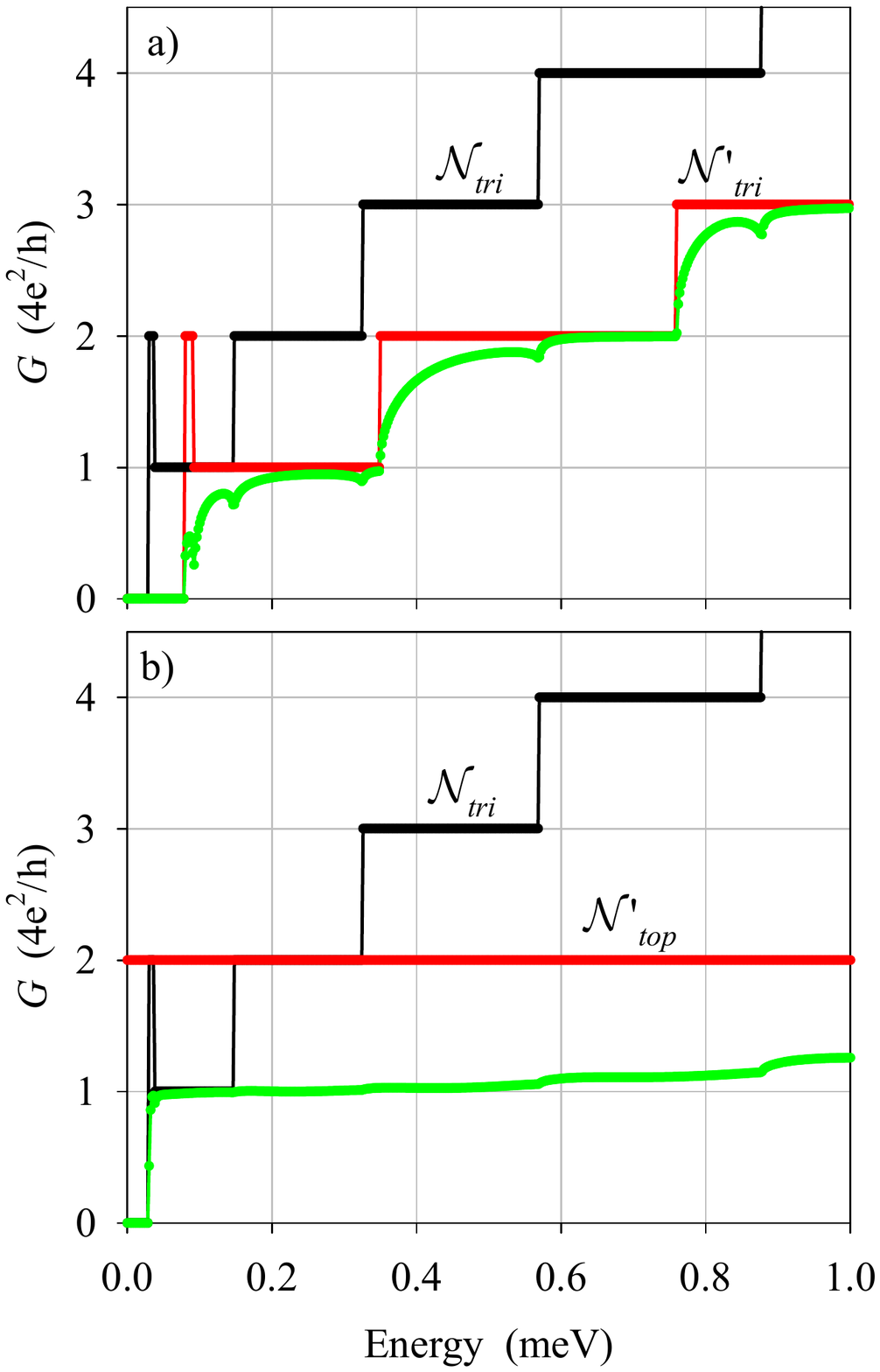}
\end{center}
\caption{Conductance of electrostatic single junctions (green)
of trivial-trivial type (a), and trivial-topological type (b). 
Black and red data show the number of 
propagating modes (Chern number) to
the left and right sides of the junction,
labeled as 
${\cal N}_{\it tri}$, ${\cal N}'_{\it tri}$ (a)
and 
${\cal N}_{\it tri}$, ${\cal N}'_{\it top}$ (b), respectively.
Parameters: $L_y=600$ nm,  $L'_y=400$ nm, $|V_a|=20$ meV, $s=12$ nm.}
\label{Fig2}
\end{figure}

Figure \ref{Fig2} compares, as a function of energy, the conductances of the 
trivial-trivial (\ref{Fig2}a) and trivial-topological (\ref{Fig2}b) 
junctions sketched in Fig.\ \ref{Fig1}. 
Both panels show the same 
staircase evolution 
of the number of propagating 
modes on the left, ${\cal N}_{\it tri}$ (black), but the number of modes 
on the right, ${\cal N}'_{\it tri}$ or ${\cal N}'_{\it top}$, differs. 
Figure \ref{Fig2}a shows that
${\cal N}'_{\it tri}$ also presents 
a similar staircase behavior as ${\cal N}_{\it tri}$,
while in Fig.\ \ref{Fig2}b 
${\cal N}'_{\it top}$ remains at a constant value 2.
The behavior of  ${\cal N}_{\it tri}$ and 
${\cal N}'_{\it tri}$ are as expected, 
since the band structure of the trivial wire is such that 
successive modes are activated as the Fermi energy overcomes successive band
minima. On the other hand, the topological wire contains two branches $E(k)$ without 
a corresponding band minimum, but crossing from negative to positive energies 
with a fixed slope.\cite{Mar08}

The conductance of the trivial-trivial junction (green, Fig.\ \ref{Fig2}a)
saturates to ${\cal N}'_{\it tri}$ as the energy is increased, with
the modifications of
a rounding of the ${\cal N}'_{\it tri}$ steps and small conductance dips at
the onset of the ${\cal N}_{\it tri}$ steps. 
These features are rather similar to the results of
semiconductor wire junctions in 2DEGS, the smoothened  conductance being due to 
wave function reflections at the junction interface for energies near 
the activation onset of propagating modes. 

In sharp contrast, the conductance of the
trivial-topological junction (green, Fig.\ \ref{Fig2}b) displays a 
conspicuously different behavior. In this case the conductance does not 
saturate to ${\cal N}'_{\it top}$, at least not with a fast convergence 
as in the upper pannel. Actually, $G$ settles to a value close to 1 (in units
of $G_0=4e^2/h$), in a plateau-like behaviour, with deviations of the quantized value
becoming visible only when ${\cal N}_{\it tri} > 4$.

The conductance quench 
of Fig.\ \ref{Fig2}b
to a value $G\approx 4e^2/h$, even when the topological kink-antikink
system could in principle conduct ${\cal N}'_{\it top}=2$ propagating modes 
(per valley and spin), is a remarkable feature. It is indicating that 
the incident modes from the trivial-side are mostly reflected except 
for one mode which is transmitted. We know that 
the modes on the topological side (kink-antikink)
are localized to the $y$ values near the kink or the antikink  
and they have a valley dependent chirality. That is,  
two $K$ modes can be transmitted in the lower ($y<0$) kink, while 
two $K'$ modes can be transmitted in the upper ($y>0$) antikink.
A priori, the modes on the left (trivial) side are expected to be 
nonchiral, mostly propagating along the bulk of the wire $-L_y/2<y<L_y/2$.
The observation that one mode is transmitted to the right, however, is strongly 
suggesting that one particular mode of the trivial wire is also 
edge-like and 
chiral, so that it can effectively couple to the chiral modes 
on the right side.

\begin{figure}[t]
\begin{center}
\includegraphics[width=0.5\textwidth, trim = 1cm 7.5cm 2cm 4cm, clip]{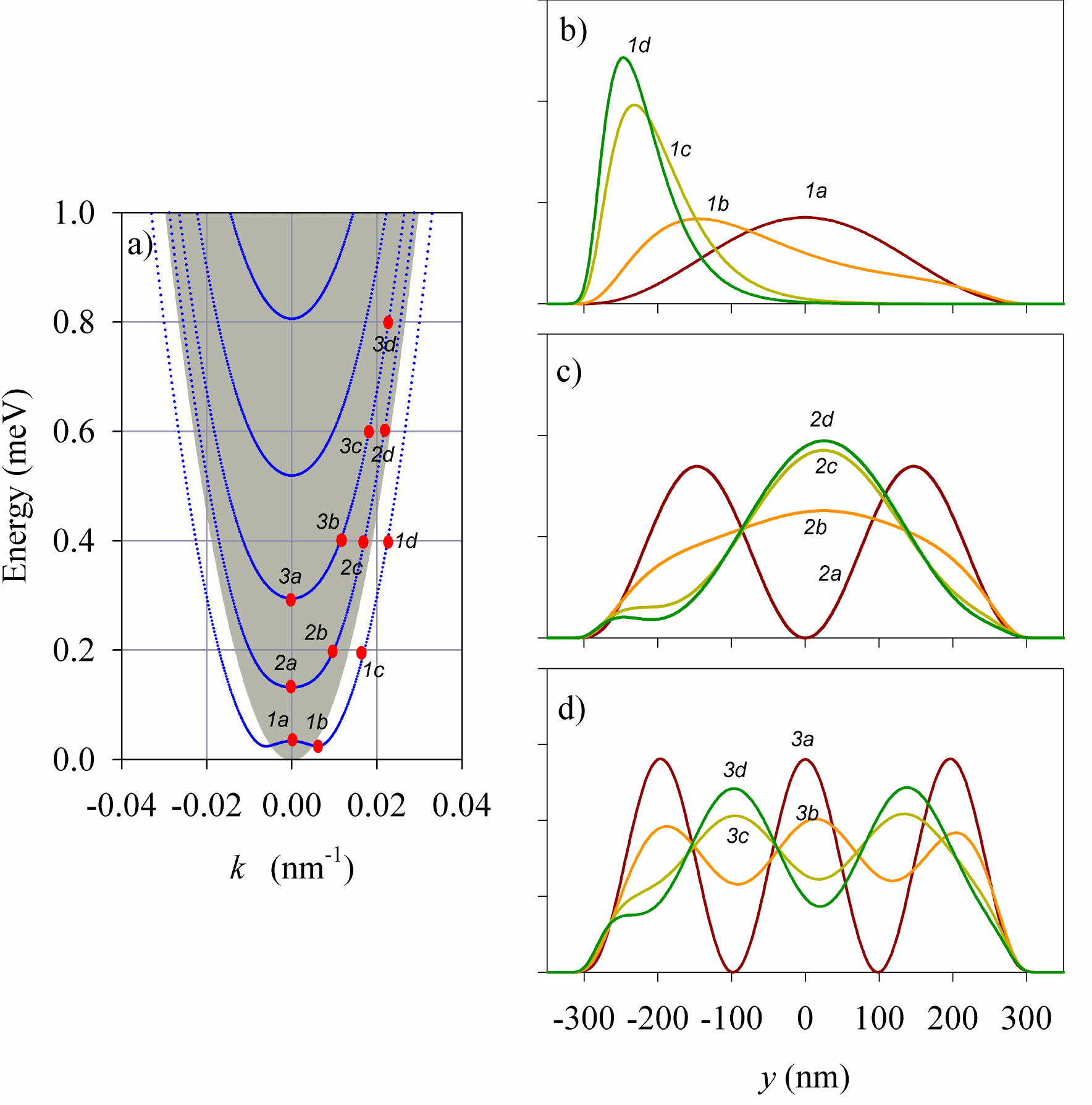}
\end{center}
\caption{ a) Energy bands of the trivial confinement wire with $L_y=600$ nm, $|V_a|=20$ meV, $s=12$ nm. 
The shaded region corresponds to 
the existence of real-$q$ modes,
$k<k_c$ with $k_c$ the critical value given in Eq.\ (\ref{eq9}).
b,c,d) Probability densities for the $k$ values of the successive bands indicated with red dots in panel a).}
\label{Fig3}
\end{figure}

The emergence of an edge chiral mode in an electrostatic BLG wire of trivial
confinement is a surprising result, more so in absence of magnetic field
such as our case. 
It is the main finding of our work.
A valley-momentum-locked edge mode does not necessarily require
a topological confinement with two nearby electrodes of opposite signs,   
as in the right side of Fig.\ \ref{Fig1}b, but the 
simpler trivial electrode distribution in the left side of Fig.\ \ref{Fig1}b or in Fig.\ \ref{Fig1}a is also enough to sustain such a mode at the border between biased and unbiased BLG regions. 
This rather counterintuitive result is further discussed in the remaining of this Section and in  
Secs.\ \ref{s3b} and \ref{s3c}.

Next, we validate our above interpretation by analysing the 
physical character of the modes of a trivial wire with numerical and 
analytical results.
Figure \ref{Fig3} shows the band structure of a trivial 600-nm-wide wire
and the spatial distributions of probablity densities for a few selected states
along
the lower energy branches. It only displays results for positive energy states
since the negative-energy ones are mirror symmetric by energy and $y$ inversion,
$E\to -E$ and $y\to -y$. Notice also that, while the energy branches are valley 
degenerate, the shown spatial distributions correspond to valley $K$. The
corresponding distributions for $K'$ are again given by $y$ inversion.

The results of Fig.\ \ref{Fig3} prove that the lowest branch 
of a trivial wire indeed becomes edge and chiral as 
the wave number $k\equiv p_x/\hbar$ is increased 
along the branch. Those results have been obtained numerically but is 
also possible to perform an analytical analysis. 

\subsection{Analytics}
\label{s3b}

Assuming 
wave numbers $k$ and $q$ along $x$ and $y$, respectively, and a constant $V_a$
in Eq.\ (\ref{eq1}) the eigenmode equation becomes an algebraic  
$4\times 4$ matrix problem for each valley, i.e., replacing
$\tau_z\to s_\tau$ with $s_\tau=\pm 1$ for valley $K\,(K')$. 

In the purely homogenous case we can assume a
spinorial wave function ($\sigma,\lambda=1,2$)
\begin{equation}
\label{eq3}
\Psi \equiv \Phi_{\sigma\lambda}\, e^{i(kx+qy)}\; ,
\end{equation}
with a 4-component spinor of constants $\Phi_{\sigma\lambda}$
for each $k$ and $q$. 
These constants are determined from the eigenvalue problem 
$H\Psi = E\Psi$ with the Hamiltonian of Eq.\ (\ref{eq1}).
We rewrite the eigenvalue problem as
\begin{equation}
\sigma_y H \Psi = E \sigma_y \Psi\; ,
\end{equation}
and with the $\Psi$ of Eq.\ (\ref{eq3}) it can be recast as
\begin{eqnarray}
\frac{1}{\hbar v_F}
\left[
\rule{0cm}{0.5cm}
E\,\sigma_y +
i s_\tau\, \hbar v_F\,  k\, \sigma_z \right.
&+&
\frac{t}{2}\left(
i\,\lambda_x\sigma_z -\lambda_y
\right) \nonumber\\
&-& \left. V_a\, \sigma_y \lambda_z 
\rule{0cm}{0.5cm}
\right]
\Phi
=
q\, \Phi\; .
\label{eq5}
\end{eqnarray}

Equation (\ref{eq5}) is a $4\times 4$ eigenvalue problem, 
${\mathcal M}\,\Phi=q\,\Phi$, determining the transverse  wave numbers
$q$
for a given $E$ and $k$ from 
\begin{equation}
\det \left(
{\mathcal M}- q\, \mathbf{1}
\right) = 0\; .
\end{equation}
Straightforward algebra yields
\begin{eqnarray}
q = 
\pm
\frac{1}{\hbar v_F}
\left[
\rule{0cm}{0.4cm}
\right.
&-& \hbar^2\, v_F^2\, k^2 
+ V_a^2
+E^2 \nonumber\\
&\pm &
\sqrt{
4 V_a^2 E^2 +t^2(E^2-V_a^2)
}\,
\left.
\rule{0cm}{0.4cm}
\right]^{1/2}\; .
\label{eq7}
\end{eqnarray}

Equation (\ref{eq7}) yields four $q$ roots whose character as
purely real or complex numbers determine whether states having propagating or
evanescent character along $y$ can emerge. 
If, for given $E$ and $k$,
Eq.\ (\ref{eq7}) has no real $q$, it necessarily implies that 
only transverse decaying states can emerge for those $E$ and $k$ values.
In a large $V_a$, either positive or negative, all $q$'s are complex 
for reasonable $E$ and $k$ values, 
and thus 
states must necessarily 
decay. This is not surprising, since a large $V_a$ causes the decay
in the sides of a trivial wire (Fig.\ \ref{Fig1}a), and also the decay
in the two directions when $y$ departs from the kink and  antikink 
positions (Fig.\ \ref{Fig1}b). Notice that a negative $V_a$ causes
a similar decay of a positive $V_a$, a usual property of
relativistic-like Dirac systems.

A surprising behavior 
for unbiased ($V_a=0$)  BLG 
is that Eq.\ (\ref{eq7}) still predicts a range of $E$ and $k$ values 
such that all $q$'s are complex. Naively, one could expect that unbiased
BLG, being gapless, would only sustain bulk propagating states for any 
$k$. However, Eq.\ (\ref{eq7}) for $V_a=0$ reads
\begin{equation}
q=
\pm
\frac{1}{\hbar v_F}
\sqrt{
-\hbar^2 v_F^2 \, k^2 + |E| \left(\, |E| \pm t\, \right)
} \; ,
\label{eq8}
\end{equation}
and, since $|E| < t$, it is then clear from Eq.\ (\ref{eq8}) that all
$q$'s are purely imaginary for $k>k_c$ where $k_c$ is the critical value 
\begin{equation}
k_c = \frac{1}{\hbar v_F}\, \sqrt{ |E|\, \left(\, |E| + t\, \right)}\; .
\label{eq9}
\end{equation}

The shaded area in Fig.\ \ref{Fig3}a corresponds to $k<k_c$, where real $q$'s 
exist. Notice that the lowest branch 
of Fig.\ \ref{Fig3}a
is then in the region of transverse 
decaying states, except for $k$ close to zero where the band presents a 
small maximum. This confirms the emerging edge chiral character of the 
lowest branch as $k$ is increased seen in Fig.\ \ref{Fig3}b, as well as
the bulk character of the states in Figs.\ \ref{Fig3}cd.

\begin{figure}[t]
\begin{center}
\includegraphics[width=0.5\textwidth, trim = 4cm 1cm 4cm 2cm, clip]{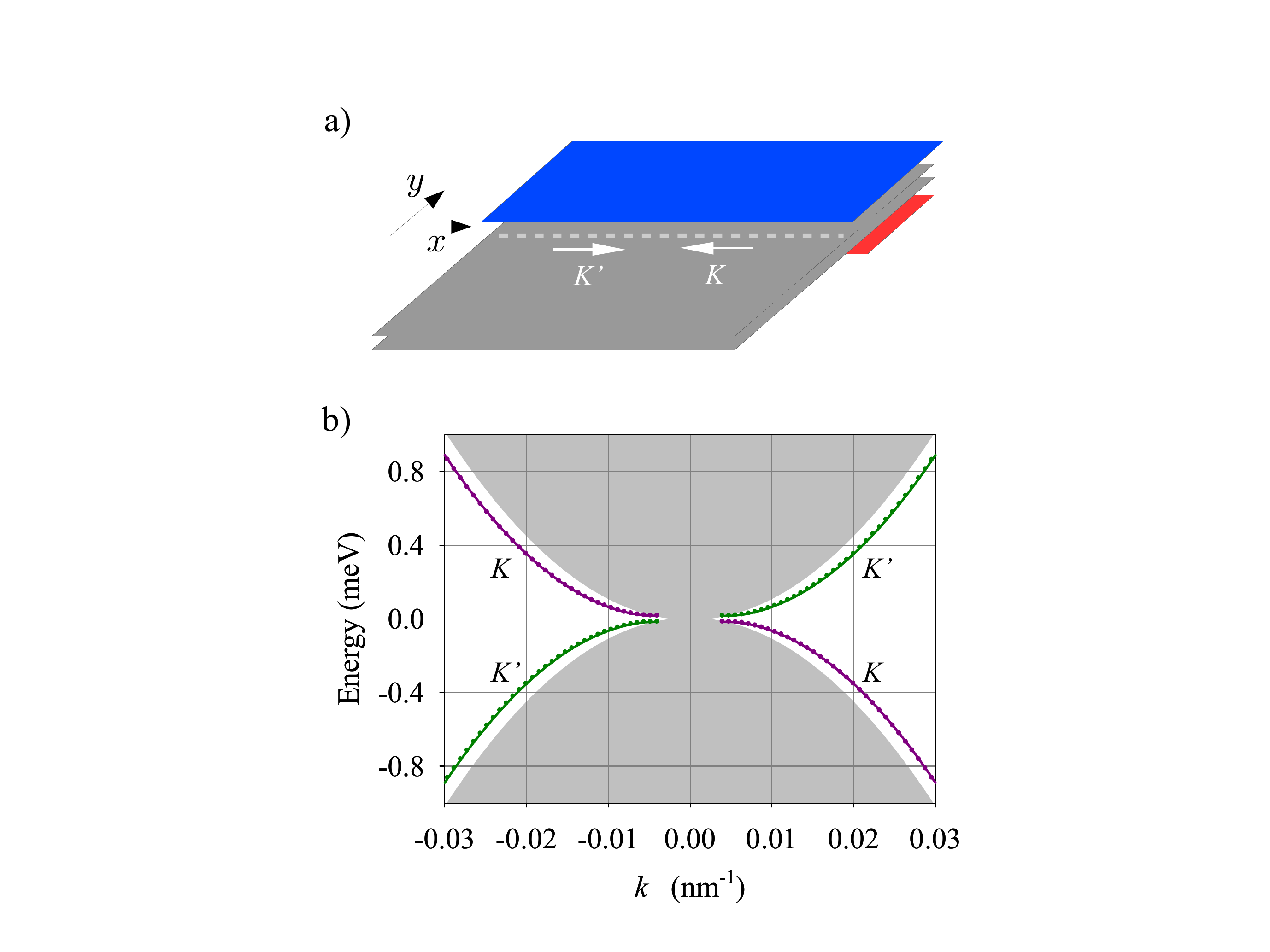}
\end{center}
\caption{
a) Sketch of a semi infinite interface between unbiased and biased 
BLG. The arrows indicate the propagation of the edge chiral mode
for $K$ and $K'$ valleys along the interface  shown with a dotted line.
b) Energy dispersion of the edge chiral modes. The shading corresponds 
to the continuum of states for $k<k_c$ of Eq.\ (\ref{eq9}). Parameters: 
$|V_a|=20$ meV, $s=12$ nm. 
}
\label{FigTT}
\end{figure}

\subsection{Semi infinite unbiased-biased interface}
\label{s3c}

Having identified an edge chiral mode in a trivial wire from the conductance
of a trivial-topological junction, a natural question to address next 
is whether this mode is also present  or not in a semi infinite interface
between unbiased and biased BLG (Fig.\ \ref{FigTT}a). We have then calculated 
the transverse localized states by numerically imposing zero boundary condition for
$y\to\pm\infty$ with the electrode configuration sketched in Fig.\ \ref{FigTT}a.
Indeed, for $k>k_c$ a localized state is found with the energy dispersion 
of Fig.\ \ref{FigTT}b for $K$ (magenta) and $K'$ (green) valleys.

Figure \ref{FigTT} highlights the opposite chiralities of the $K$ and $K'$ 
edge modes propagating along the dotted line interface. The modes are characterized by
their opposite slopes for $K$ and $K'$, indicating 
their valley-momentum locking, similarly to the topological modes of  
a kink.\cite{Mar08}
However, clear differences with the kink modes are:
a) only one mode
per valley is present in Fig.\ \ref{FigTT} 
while there are two kink modes per valley; 
b) kink modes
lie in an energy-gapped region of the spectrum, while the present edge chiral 
modes co-exist with bulk modes of the shaded area of Fig.\ \ref{FigTT} 
lying at the same energy. In fact, the edge chiral branches even merge with the continuum
for $k$ close to zero. Appendix \ref{append} provides further analysis 
of the edge chiral modes
using quasi-analytic complementary approaches and also  
investigating their robustness against 
the diffusivity $s$ and the value of the asymmetry potential $V_a$.

\begin{figure}[t]
\begin{center}
\includegraphics[width=0.5\textwidth, trim = 3.8cm 0.5cm 2.2cm 1.3cm, clip]{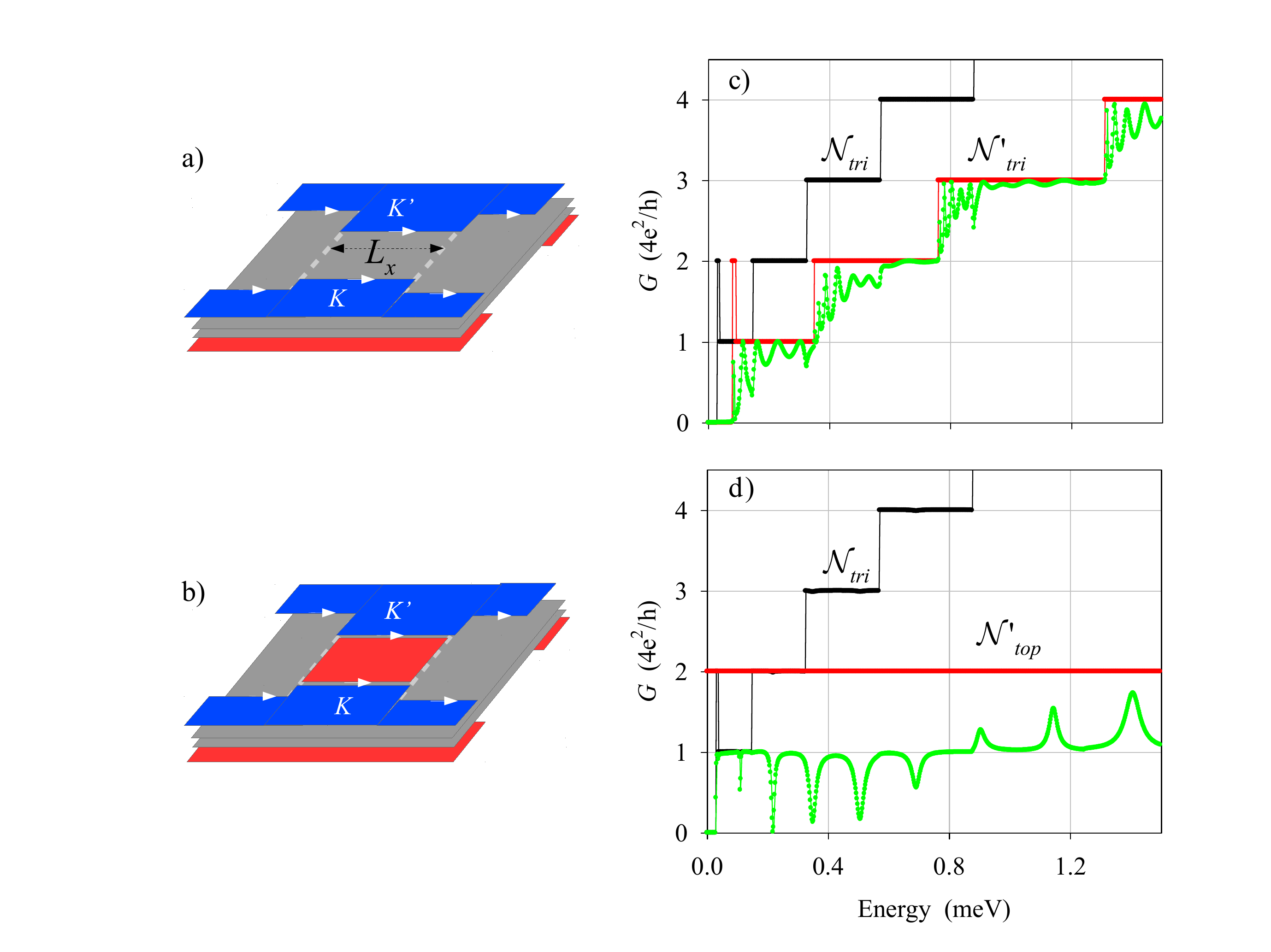}
\end{center}
\caption{
a,b) Sketches of the double junction setups 
similar to Fig. \ref{Fig1}.
The length of the central part is $L_x=1\;\mu{\rm m}$. 
c,d) Results of the double junctions shown on the corresponding left panels. 
Black line is indicating the number of modes in trivial left and right leads, 
while red is the number of modes in the center, which can be either  trivial (a,c) or topological (b,d).
Rest of parameters as in Fig.\ \ref{Fig2}.
}
\label{Fig4}
\end{figure}

\subsection{Wire double junctions}

As a final item of this work, we address the study of the 
double junctions with trivial leads and a scattering center which is either 
also trivial or topological. Sketches of the two types of double junctions
can be seen in Fig. \ref{Fig4}ab. In these geometries a new parameter 
appears as compared to the preceding 
single junctions, the length $L_x$ of the central section. The double junction 
with a trivial center (Fig.\ \ref{Fig4}c) displays similar results 
to the trivial-trivial single junction (Fig.\ \ref{Fig2}a). The conductance
saturates to ${\cal N}'_{\it tri}$ as the energy increases. The rounded steps 
are now transformed into short oscillations. They are caused by interferences
of Fabry-P\'erot type due to multiple reflections between the first and the second interfaces of the junction.
The oscillation amplitude is larger near the activation thresholds of
transmitted modes, ${\cal N'}_{\it tri}\to{\cal N'}_{\it tri}+1$, and
it smoothly decays for increasing energy until the next activation threshold is reached. The thresholds for incident modes, ${\cal N}_{\it tri}\to{\cal N}_{\it tri}+1$, also
leave a trace on the conductance curve and, in particular, 
a rather flat conductance is seen for 
$({\cal N}_{\it tri},{\cal N'}_{\it tri})=(4,2)$ and
$(5,3)$.
The Fabry-P\'erot interferences would be strongly enhanced
in presence of barriers at the junction interfaces (dotted lines in Fig.\ \ref{Fig4}a) but this would require using additional electrodes.

The conductance of the double junction with a topological center 
(Fig.\ \ref{Fig4}d) shows outstanding features. 
The quenching to $G\approx 4e^2/h$ due to the transmission of a single 
edge chiral mode of the trivial wire is again observed, as in Fig.\ \ref{Fig2}b.
However, a remarkable sequence of dips and peaks is now found
in the double junction. These features are an evidence of the 
spectrum of topological eigenstates forming closed loops around the central
region of the double junction. The energies of those topological loops can be 
well described by a Bohr-Sommerfeld quantization rule,\cite{Ben21b}
requiring an integer number of wave lengths fit into the loop perimeter.
The dips or peaks are then consequences of Fano resonances due to the coupling 
of localized states, the closed loops, with the scattering states. The effective 
coupling varies with energy and transforms the resonances from dips 
at low energy into peaks at higher energy
in Fig.\ \ref{Fig4}d. 

Fano resonance profiles are characterized as\cite{Fano61,Nock94,Sanc06}
\begin{equation}
G \approx \frac{ | \epsilon + q |^2}{\epsilon^2 + 1}\; , 
\end{equation}
where $q$ is the Fano parameter and the energy dependence is 
contained in $\epsilon\equiv (E-E_R)/\Gamma$, with $E_R$ and $\Gamma$ the resonance energy and width. 
Different values of the Fano parameter $q$ 
yield varying resonance profiles, from symmetric dips ($q\to0$)
to intermediate asymmetric profiles (finite $q$'s) and symmetric peaks (large $q$'s).
Figure \ref{Fig4} is indicating a fast evolution of the Fano 
parameter with energy, asymmetry profile being observed only
around 0.85 meV for that specific set of parameters.
A detailed analytical modeling of the coupling between the 
topological closed loops
and  the scattering states 
requires a fine tuning of the coupling intensities
and is out of the scope of the present work.

\section{Conclusions}

Junctions of electrostatically confined BLG wires show outstanding 
transport features feasible to be detected.
A trivial-topological junction is characterized by a conductance quench
with respect to a trivial-trivial junction. A single mode of edge chiral 
character, per valley and spin, is transmitted from the trivial to the 
topological side causing a near quantization $G\approx 4e^2/h$ when the number 
of incident modes is low (${\cal N}_{\it tri}< 4)$. With larger 
values of ${\cal N}_{\it tri}$ the conductance deviation from the quantized value
becomes increasingly visible.

The edge chiral character of 
the lowest mode of a trivial BLG wire is surprising, 
more so in the absence of any 
magnetic field. Such mode is also present at a semi infinite interface 
between unbiased and biased BLG planes. It is characterized by 
a locking between valley and momentum and there is an analytical  
critical value of momentum
$k_c$ for the presence of such mode when $k>k_c$, 
which is otherwise damped in the continuum
of bulk states. 

Double junctions of electrostatic BLG wires with 
trivial leads and 
a topological center
show conspicuous Fano resonances, dips or peaks in the conductance. 
They are due to an energy-dependent coupling of the closed loop 
around the topological center and the scattering states. Altogether, 
the above features may help to place 
hybrid trivial-topological
BLG graphene wires with electrostatic confinement
as a useful and
controllable platform for graphene electronics.

\acknowledgments
Helpful discussions with David S\'anchez
are gratefully acknowledged.
We acknowledge support from Grant
No. MDM2017-0711
and Grant
No.\ PID2020-117347GB-I00
funded by MCIN/AEI/10.13039/501100011033,
and Grant
No. PDR2020-12
funded by 
GOIB.

\appendix

\section{The edge chiral mode}
\label{append}

In this appendix we analyze further the existence of an edge chiral mode
at the boundary between unbiased ($V_a=0$) and biased ($V_a\ne0$) BLG
as sketched in Fig.\ \ref{FigTT}a. Specifically, we obtain such mode in two complementary approaches.
First  
we use a $y$-grid diagonalization as 
in Fig.\ \ref{FigTT}b
to investigate the dependence of the mode dispersion on 
the asymmetry potential diffusivity $s$ and on its 
saturating value
$V_a$. 
Second, we use a quasi-analytic method valid 
for the sharp interface, imposing the matching of the properly decaying 
solutions on the two sides of the interface. The two 
approaches are in excellent agreement and thus support the physical existence 
of the edge chiral mode.

\begin{figure}[t]
\begin{center}
\includegraphics[width=0.45\textwidth, trim = 2.5cm 5cm 2.5cm 1cm, clip]{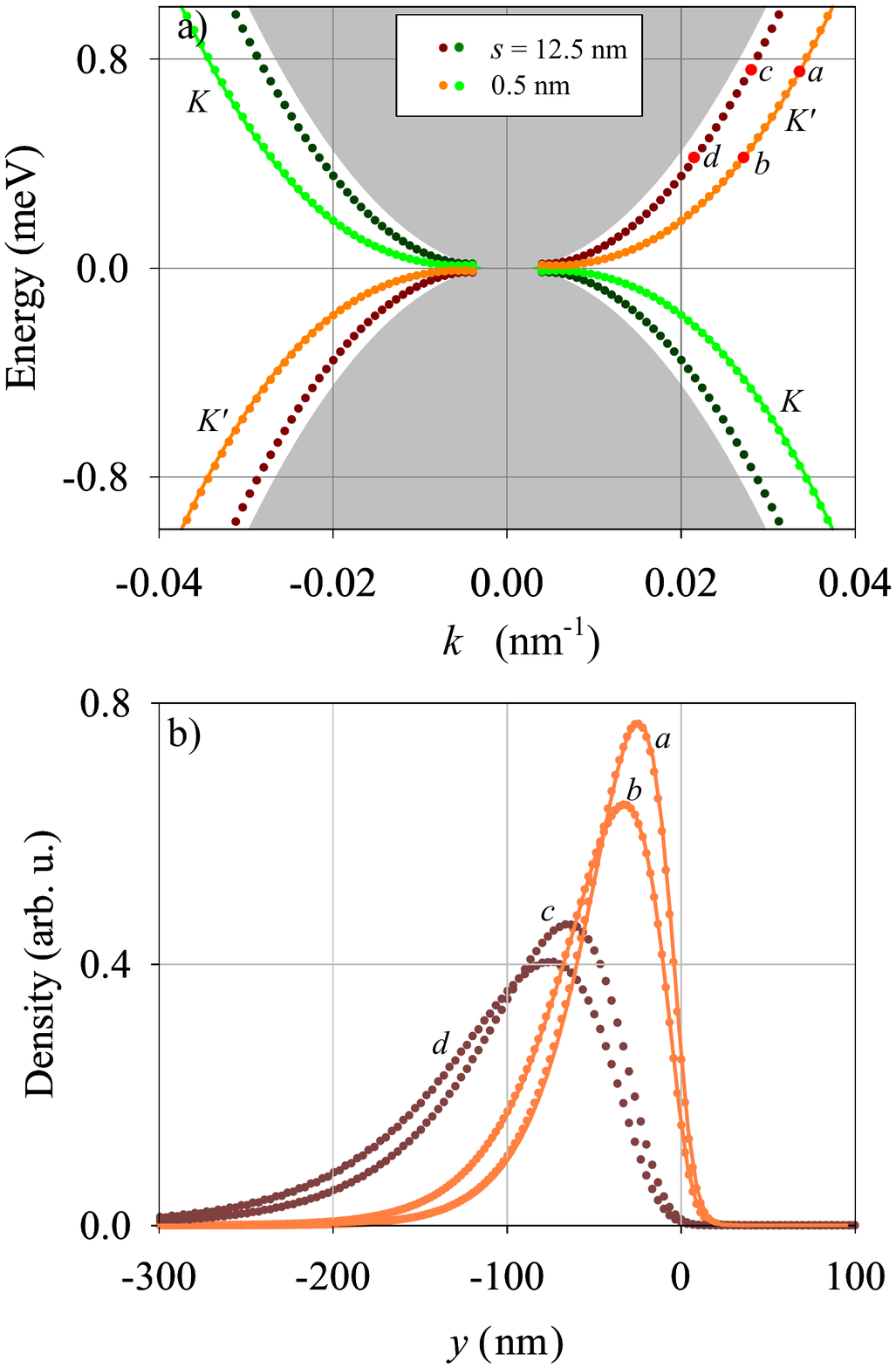}
\end{center}
\caption{
a) Energy dispersion of the $K$ and $K'$ edge chiral mode for the 
semi infinite interface (Fig.\ \ref{FigTT}a)
and for two different values of the diffusivity parameter $s$. 
We have assumed $V_a=20$ meV and the shaded region corresponds to $k<k_c$.
b) Spatial distribution of the probability densities for the indicated states of panel a). 
The solid lines in panels a) and b) are the results 
of the quasi-analytic model for a sharp interface
and they show an excellent agreement with 
the $s=0.5$ nm smooth potential data.
}
\label{Figap}
\end{figure}

Figure \ref{Figap} addresses the dependence on the smoothness $s$, 
showing that 
the sharper the asymmetry potential the larger 
the separation of the mode branch from the continuum indicated in gray.
The lower panel Fig.\ \ref{Figap}b shows that the probability densities 
with a larger $s$ extend farther from the interface than the densities 
for smaller $s$, as one could intuitively expect. We have also obtained (not shown in the figure) that the probability distributions 
are unchanged when reversing the momentum ($k\leftrightarrow -k$) for a given valley, or
reversing the valley ($K\leftrightarrow K'$) for a given momentum.

In case of a sharp transition interface $s=0$ we have solved the matching 
condition at the interface  considering exponentially decaying solutions  
towards both sides. The $q$ wave numbers
and $\Phi$ wave functions are determined from 
Eq.\ (\ref{eq7}) and Eq.\ (\ref{eq5}), respectively.
The matching approach does not use any grid discretization
of the $y$ coordinate and only requires the 
calculation of $4\times 4$ matrices,  as determined by the number of $q$ modes. 
The results are in excellent agreement 
with the grid result for small diffusivity, as shown by the solid lines of Fig.\ \ref{Figap}a, thus  
confirming the physical character of the edge chiral mode. 
We also stress that the two methods 
prove that there is only one branch for $K$ and one branch
for $K'$, differently to the result for kink states yielding two
branches per valley.\cite{Mar08,Zarenia11} 
Notice, however, that the present chiral edge branches are not topologically protected  from the continuum of bulk states (gray) by an energy gap, as occurs for the kink states.

\begin{figure}[t]
\begin{center}
\includegraphics[width=0.45\textwidth, trim = 2.3cm 2cm 2.5cm 3.5cm, clip]{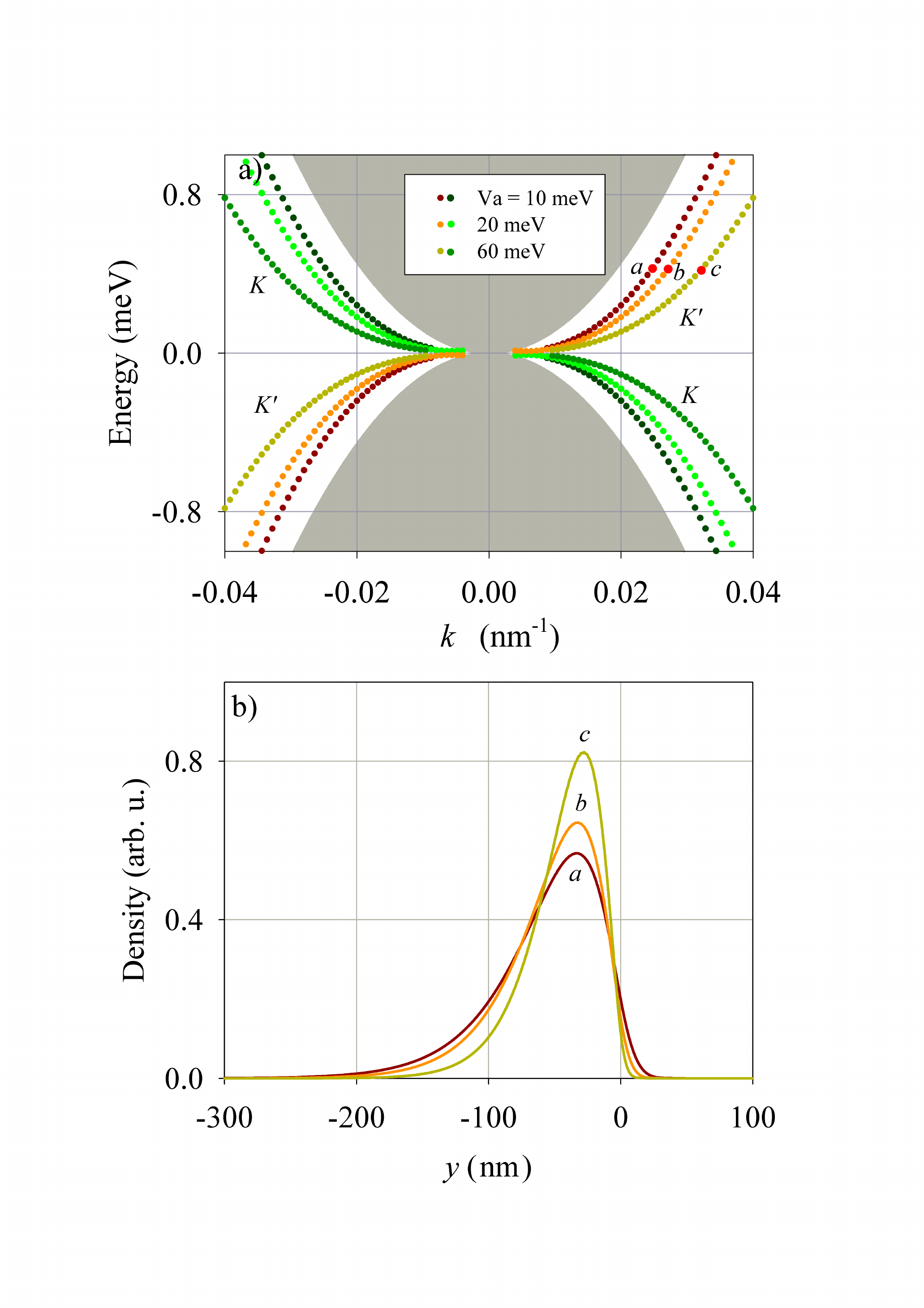}
\end{center}
\caption{Same as Fig.\ \ref{Figap} for different values of the 
asymmetry potential $V_a$. We assumed a sharp interface $s=0.5$ nm.}
\label{Figap2}
\end{figure}

Finally, the dependence on $V_a$ is analyzed in Fig.\ \ref{Figap2}.
The energy dispersion $E(k)$ becomes flatter with increasing $V_a$
(Fig.\ \ref{Figap2}a) while the density distribution becomes narrower
(Fig.\ \ref{Figap2}b), corresponding to a more localized state at the interface. Nevertheless, the mode features 
are qualitatively preserved even with large changes in $V_a$.

\bibliography{BGJbib}

\end{document}